\documentstyle[11pt,newpasp,twoside,epsf]{article}

\def\edcomment#1{\iffalse\marginpar{\raggedright\sl#1\/}\else\relax\fi}
\reversemarginpar

\begin{document}
\title{Obstacles to the Collisional Growth of Planetesimals}
 \author{Andrew N. Youdin}
\affil{Princeton University Observatory, Princeton, NJ 08544}
\begin{abstract}
The debate over whether kilometer-sized solids, or planetesimals, 
assemble by collision-induced chemical sticking or by gravity-driven unstable modes remains unsettled.   In light of recent work showing that gravitational growth can occur despite turbulent stirring, we critically evaluate the collisional hypothesis.  Impact speeds in protoplanetary disks reach $\sim 50$ m/s in a laminar disk and may be larger in turbulent disks.  We consider the role of elastic and plastic deformations, restructuring of ``rubble piles," and sticky organic matter.  Coagulation is possible for dust grains with large surface area-to-volume ratios.  Simple energetic arguments show that bouncing, cratering, and fragmentation should dominate collisional dynamics between millimeter and kilometer sizes.
\end{abstract}

\section{Introduction}\label{intro}
Solid planets and gas giant cores form in a hierarchical growth process spanning some 13 orders of magnitude in radius from interstellar dust grains to spectacular giant impacts.  An important demarcation occurs at kilometer sizes, where coupling to the gas disk is weak and substantial escape speeds of $\approx 1$ m/s allow growth by inelastic binary collisions.  Two theories aim to explain the origin of planetesimals from smaller solids with negligible surface gravity and strong gas couplings.
This paper examines what we will call the Collisional-Chemical Sticking Hypothesis (hereafter CCSH), that km-sized bodies arise from a long chain of agglomerative growth, due to chemical forces during inelastic collisions.
The gravitational instability (hereafter GI) hypothesis appeals the collective self-gravity of a dense midplane layer to collect small solids into larger planetesimals (Goldreich \& Ward 1973, Youdin \& Shu 2002, hereafter YS).  The characteristic wavelength of GI predicts planetesimal sizes $\ga$ few km for disks with enough mass to produce a solar system like our own.  Leapfrogging intermediate sizes by direct assembly from $\la$ cm solids avoids most problems associated with sticking and rapid radial drift (see \S\ref{laminar}). 

The main obstacle to GI is that turbulence opposes the settling required to create a dense midplane layer.  Even an otherwise laminar disk  generates midplane turbulence, since particle settling introduces destabilizing vertical shear (Weidenschilling \& Cuzzi 1993, hereafter WC).  Sekiya (1998) overcame this obstacle by appealing to the enhanced buoyancy of stratified, high metallicity disks.  
 YS showed that turbulent motions can only suspend a finite surface density of solids, $\Sigma_{\rm p,crit}$.  GI occurs in a dense midplane layer when the actual $\Sigma_{\rm p}$ exceeds this saturation threshold.  However the required surface density ratio of solids to gas, $\Sigma_{\rm p,crit} / \Sigma_{\rm g}$, i.e.\  the disk metallicity, is about an order of magnitude over solar abundances.
   Various enrichment mechanisms were described in YS.  A particularly robust finding is that aerodynamic radial drift has the global consequence of enhancing $\Sigma_{\rm p}$ as the particle disk shrinks in radius.  Youdin \& Chiang (2003) explored these ``particle pileups" in detail,  including inertial effects and 
self-consistently generated turbulent stresses in the particle layer.

Since GI can overcome the heretofore fundamental obstacle of vertical shear, we assess the problems facing the CCSH.  YS argue against the plausibility of the CCSH with evidence based on micro-gravity experiments, the size distribution of meteoritic inclusions, the limited sticking ability of icy planetary ring debris, and ``intuition."  Our goal here is to physically explain the ``intuitive" notion that collisions between macroscopic solids with weak surface gravity, at speeds characteristic of freeway driving, are not cohesive.

The most fundamental obstacle to sticking is energetic: the \emph{entire} relative kinetic energy of two bodies at large distances, $K_\infty = m v_\infty^2/2$, must be dissipated, where $m$ is the reduced mass and $v_\infty$ is the relative speed at large distances.  This requires some combination of (a) efficient energy loss, and (b) binding energies, $U_{\rm bind} < 0$, which increase kinetic energy on impact: $K_{\rm imp} = K_\infty + |U_{\rm bind}|$.  The dissipative efficiency needed for sticking, $K_\infty/K_{\rm imp} = 1/(1+ \theta)$, drops appreciably below unity when the focusing factor $\theta \equiv |U_{\rm bind}|/K_\infty \ga 1$.  Gravitational focusing, which scales as $\theta \propto R^2/v_\infty^2$, affects bodies with sufficiently large radius, $R$.  For small bodies, surface energies, such as van der Waals interactions, give $\theta \propto R^w/v^2$, where the size falloff ranges from $w = -1$, for a maximal contact area $\sim R^2$,  to $w = -5/3$, for elastic compression (see \S\ref{elast}).  Over a large size range, extending from mm ---  10~m, no relevant binding energy exists.  Until an incredibly efficient, yet non-destructive, dissipation mechanism is identified, the CCSH will lack a solid physical foundation.

This proceeding is organized as follows: \S\ref{speed} investigates collisional speeds expected from laminar flow (\S\ref{laminar}) and turbulent fluctuations (\S\ref{turb}). Section \ref{cm} explores collisional binding mechanisms including elastic (\S\ref{elast}) and plastic (\S\ref{plastic}) deformation as well as restructuring within ``rubble piles" and  the role of  organic matter (\S\ref{agg}).  Section \ref{concl} contains concluding remarks.

\section{Collision Speeds}\label{speed}
Before planetesimals form, gravitational scattering is negligible.  Particle random velocities arise from frictional coupling to the gas, including kicks from turbulent fluctuations, drift induced by the gas disk's sub-Keplerian rotation, and vertical settling, which becomes small near the midplane as vertical gravity vanishes.  

\begin{figure}[htb]
\plotfiddle{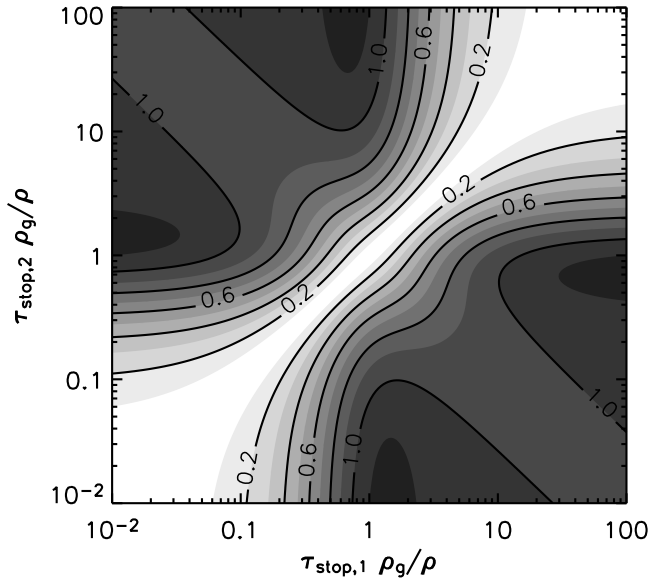}{2.9in}{0}{40}{40}{-150}{-10}
\caption{Contours of in-plane relative velocity induced by laminar gas drag, labeled in units of $\eta v_K \rho/\rho_{\rm g} \approx 54\rho/\rho_{\rm g}~{\rm m/s}$, between particles with stopping times $\tau_{\rm stop,1}$ and  $\tau_{\rm stop,2}$.
 }
\label{drift}
\end{figure}

\subsection{Laminar Drift}\label{laminar}

Aerodynamic gas drag gives particles radial, $v_r$, and azimuthal, $v_\phi$, deviations from Keplerian circular speeds, $v_K = \Omega r$ (Nakagawa, Sekiya, \& Hayashi 1986):
\begin{eqnarray}
v_r &=& -{2 \tau_{\rm stop} \eta v_K \over (\rho/\rho_g)^2 +  \tau_{\rm stop}^2},\\
v_\phi &=& -{\eta v_K  \over \rho/\rho_{\rm g} + \rho_{\rm g}\tau_{\rm stop}^2/\rho},
\end{eqnarray}
where $r$ is the disk radius, $\Omega$ the Keplerian frequency, $\eta \equiv -\partial P/\partial r/(2\rho_{\rm g}\Omega^2 r)$ measures centrifugal pressure ($P$) support; $\rho \equiv \rho_{\rm p} + \rho_{\rm g}$  is the spatial mass density of solid particles, $\rho_{\rm p}$, plus gas, $\rho_{\rm g}$; and $\tau_{\rm stop} \equiv \Omega t_{\rm stop}$ measures the coupling of solids to gas, with $\tau_{\rm stop} \ll 1$ ($\gg 1$) implying strong (weak) coupling.  The stopping time, $t_{\rm stop}$,
depends on the drag law (e.g.\ Epstein or Stokes) that applies for given particle sizes and gas properties (Weidenschilling 1977). 
 
The velocity scale, $\eta v_K \approx 54 ~{\rm m/s}$, is independent of $r$ in the 
 minimum solar nebula model (see YS), hereafter MSN.
 Different temperature profiles, $T \propto r^{-q}$, give $\eta v_K \propto r^{1/2-q}$.
 Our discussion ignores the inertial scaling factor $\rho_{\rm g}/\rho < 1$, which is ${\mathcal O}(1)$ unless $\rho_{\rm p} > \rho_{\rm g}$, in which case GI should develop (YS).  Particles with $\tau_{\rm stop} \approx 1$ experience the fastest radial drift, $v_{r,\rm{max}} \approx  -\eta v_K$.  These marginally coupled particles have a size  $R_{\tau 1} \approx 1$ m, at 1 AU in the MSN, giving the well-known and problematic drift timescale: $\sim {\rm AU}/(\eta v_K) \approx 100$ yr.  Elsewhere a meter is less significant. 
At larger $r$, $\rho_{\rm g}$ decreases giving weaker coupling and smaller $R_{\tau1}$ values.  Beyond about $3$ AU, $\tau_{\rm stop} = 1$ particles lie in the Epstein regime (size less than the gas mean free path) and obey
$R_{\tau1} = \Sigma_{\rm g}/(\sqrt{2\pi} \rho_{\rm s}) \approx 10 ~{\rm cm} ({r / 10 ~{\rm AU}})^{-3/2}$.
 Note that as any particle tries to grow past the local $R_{\tau 1}$ value, to slow its radial drift, it continually moves to regions where $R_{\tau 1}$ is even larger.  This struggle could be more futile, if shorter lived, than Sisyphus's since all motion is downhill (radially inward).
 
Figure \ref{drift} (generalized from figure 3 of WC) plots the relative velocity, arising from $v_r$ and $v_\phi$ drift, 
between particles with a range of stopping times, and thus sizes.  Only similarly sized particles avoid ${\mathcal O}(\eta v_K)$ relative motions.    Impacts above a threshold speed, $v_{\rm cr}$ will not give coagulation.  While $v_{\rm cr}$ uncertain and size dependent, $v_{\rm cr} \ll 50~{\rm m/s}$ is reasonable (as discussed in \S\ref{cm}).  Thus collisional growth requires maintaining a narrow size distribution.  This is a difficult proposition since many processes; e.g.\ collisional cascades, radial migration, and vertical settling, broaden the size dispersion.

\subsection{Turbulent Fluctuations}\label{turb}
Protostellar disks must be at least intermittently turbulent for stars to accrete.  
This turbulence is often modeled as an anomalous viscosity, $\nu_{\rm T} = \alpha c_{\rm g}^2/\Omega$, where $c_{\rm g}$ is the gaseous sound speed and the uncertain viscosity parameter takes values $10^{-4} < \alpha < 10^{-2}$ to reproduce T Tauri accretion rates.  Solids couple to available turbulent eddies, augmenting the particulate velocity dispersion, $c_{\rm p}$.

Assume a Kolmogorov spectrum whose largest eddies have a turnover time $t_0 \sim 1/\Omega$, and thus a typical speed $v_0 \sim \sqrt{\alpha} c_{\rm g} \sim 30\sqrt{\alpha_{3}}(r/{\rm AU})^{-1/4}$  m/s, where $\alpha_{3}\equiv \alpha/10^{-3}$.   
 Particles with $t_{\rm stop}/t_0  \sim \tau_{\rm stop} < 1$ acquire $c_{\rm p} \sim v_0$ relative \emph{to the gas}.   Larger particles, with $\tau_{\rm stop} \gg 1$,  couple poorly to all eddies, and proceed with $c_{\rm p} \ll v_0$ on nearly Keplerian orbits.  The dispersion of particles relative \emph{to each other} is as large as $v_0$ only when one of the particles is marginally coupled, $\tau_{\rm stop} \sim 1$ (or when the two particles have $\tau_{\rm stop,1} \ll 1 \ll \tau_{\rm stop,2}$).  Pairs of small particles, both with $\tau_{\rm stop} \ll 1$,  remain so tightly coupled to eddies that they have little motion relative to each other (Markiewicz, Mizuno, \& V\"{o}lk 1991).
 
Turbulence, unlike laminar drift, imparts relative motion to particles with identical stopping times.  These motions are the largest, $\sim v_0$, near $\tau_{\rm stop} \sim 1$.  This closes the narrow ``bridge" of low speed collisions in Figure 1 (also see Figure 4 in WC), which we already noted would be difficult to cross by collisional growth.   The difficulties facing the CCSH escalate further in a turbulent disk.  

Small grains, on the other hand, can coagulate in turbulent disk, especially those with $t_{\rm stop}$ less than the eddy turnover time at the dissipation scale, $t_{\rm D} = t_0 /\sqrt{Re_0}$, where $Re_0$ is Reynolds number for the largest eddies.  This regime, where the flow is laminar and Brownian motion determines $c_{\rm p}$, corresponds to sizes, 
\begin{equation}
R < (2\pi)^{-1/4}{\Omega t_0 \over \rho_{\rm s}} \sqrt{\Sigma_{\rm g} \mu \over \sigma\alpha} \approx 100(r/{\rm AU})^{-3/4}\alpha_3^{-1/2}~\mu{\rm m},
\end{equation}
 where $\sigma$ is the atomic cross section and $\mu$ is the mean molecular weight. 

\section{Contact Mechanics}\label{cm}
To evaluate the CCSH, we need to understand the aftermath of collisions at speeds up to $v_{\infty} \sim 50$ m/s.  Do they result in coagulation (mass gain), fragmentation (mass loss), or restitution (no change in mass)?   We consider identical spheres with radius $R$ and density $\rho_{\rm s}$, unless otherwise noted.  
Collisional outcomes depend on material properties, the most relevant are probably Young's modulus, $E$; the yield strength, $Y$; and the binding energy per unit area, $\gamma$.  Other influential factors include impact parameter, non-sphericity, surface roughness, and viscoelastic creep, and are considered only briefly.

 At low impact speeds, materials deform elastically, i.e.\ reversibly.  In this regime, the ratio of applied stress (force per unit area) to induced strain (relative compression) remains constant at $E$.  When imposed stresses exceed $Y$, deformation becomes plastic, or permanent, resulting for instance in  dents.  Elastic and plastic deformation during impacts generates a contact area, $A \la \pi R^2$.  Chemical forces, mainly van der Waals interactions, generate a surface binding energy, $U_{\rm bind} = -\gamma A$, where $\gamma$ varies from $30$ --- $300$ --- $3000$ ergs/cm$^{2}$ for quartz, ice, and iron grains, respectively.

We argued in \S\ref{intro} that sticking is problematic (requires very efficient energy dissipation) for $ \theta   < 1 $.  Assuming a maximal contact area, $A \approx \pi R^2$, the upper limit on $\theta$ due to surface forces is:
\begin{equation}\label{theta}
\theta_{\rm max} = {\gamma \pi R^2 \over K_\infty} \approx 10^{-2} {\gamma_{300} \over \rho_{\rm s,3} } {{\rm cm}\over R}\left({v_{\rm rel} \over {\rm m/s}}\right)^{-2},
\end{equation}
with $\rho_{\rm s,3} \equiv \rho_{\rm s}/(3~{\rm g~cm^{-3}})$ and $\gamma_{300} = \gamma/(300~\rm{erg ~cm^{-2}})$.  This suggests that coagulation between cm-sized solids at speeds above 10 cm/s is difficult.  Moreover, surface roughness reduces effective binding forces by factors $\ga 3$ because only crests of asperities come in contact (Johnson 1985).  

\subsection{Elastic Collisions}\label{elast}
Elastic sticking is an oxymoron if one defines an elastic collision as having a coefficient of restitution, $\epsilon = 1$.  However, kinetic energy is lost during elastic deformation since elastic waves (EWs) absorb kinetic energy ($K_{\rm imp}$) during compression, but do not perfectly return deformation energy during rebound.  These wayward EWs, which one can think of as wandering from the impact region, eventually dissipate, though not necessarily during the abrupt collision. 

An elastic theory of coagulation was developed by Chokshi, Tielens \& Hollenbach 1993 (hereafter CTH) who were mainly interested in the coagulation of small ($\ll$ mm) grains for which surface binding energies are significant.  They calculate a critical approach speed, $v_{\rm cr}$, above which sticking will not occur:
\begin{equation}\label{vcr}
v_{\rm cr} \approx 4 {\gamma^{5/6} \over E^{1/3} R^{5/6} \rho_{\rm s}^{1/2}} \la 1\left({R \over {\rm mm}}\right)^{-5/6}~{\rm cm \over s}.
\end{equation}
Equation (\ref{vcr}) is physically plausible since $v_{\rm cr}$ increases with surface binding energy, $\gamma$; decreases with $R$, as the surface area-to-volume ratio falls; and decreases with Young's modulus, $E$, since stiffer materials  share smaller contact areas and place less energy into elastic waves.  The threshold speed is quite low. Equation (\ref{vcr}) gives numerical values for ice, while speeds for graphite and quartz are lowered by factors of $6.5$ and $30$, respectively.  Thus elastic effects cannot explain growth beyond mm-sizes which have drift speeds $v_r \approx 1(r/{\rm AU})^{3/2}~{\rm cm/s}$.

Nevertheless,  an order-of-magnitude ``derivation" of (\ref{vcr}) is instructive. Pressing a flattened area, $A$, on a surface with radius of curvature, $R$, requires a displacement, $\delta \sim A/R$, perpendicular to the surface.  Young's modulus relates stresses, $F/A$, imposed by a static force, $F$ to the local strain, $\sim \delta/\sqrt{A}$, giving $A \sim (FR/E)^{2/3}$ and $\delta \sim [F^2/(RE^2)]^{1/3}$.  We treat compression statically because elastic wave periods are short compared to the collisional duration.

The force that interests us is the surface binding force, which is equal and opposite to the force needed to separate two spheres, $F_{\rm sep} = -3\pi\gamma R$ (Johnson, 1985, p. 127).  This result, surprisingly independent of $E$, follows from the fact that $U_{\rm bind}$ induces a finite contact area between spheres in the absence of external forces.  Equating this surface energy, $-\gamma A$, to the work required to separate the bodies, $\sim F_{\rm sep} \delta$, and recalling  $\delta \sim A /R$, reproduces $F_{\rm sep} \sim - \gamma R$.  We can also estimate equilibrium values for $\delta \sim (\gamma^2R/E^2)^{1/3}$, $A \sim (\gamma R^2/E)^{2/3}$, and, most importantly, $U_{\rm bind} \sim  -\gamma^{5/3}R^{4/3}E^{-2/3}$.

Our heuristic argument that sticking needs $ K_\infty \la |U_{\rm bind}|$ (unless dissipation is very efficient, which should not be the case for elastic deformation) reproduces equation (\ref{vcr}) to order unity.  CTH achieved their result differently.  They estimated the amount of energy lost to EWs,  $U_{\rm lost}$, by numerically evolving a spectrum of EWs in response to time-varying collisional stresses.  They found that $U_{\rm lost}$ is a fixed fraction of the binding energy, $U_{\rm lost} \approx .4 |U_{\rm bind}|$, and $K_\infty \leq U_{\rm lost}$ gives equation (\ref{vcr}).   

The best hope for the CCSH probably involves large bodies sweeping up small grains, which dominate the relative kinetic energy and binding area.  However, the large relative velocity between unequal bodies, $\eta v_K  \approx 54$ m/s, implies via equation (\ref{vcr}) that only extremely small ``grains" could stick: 30 --- 3 --- 0.5 nm for ice, graphite, and quartz grains, respectively.  However, impact speeds might be reduced below $54$ m/s as the gas flow is deflected around the larger body.  Nevertheless, maintaining a reservoir  of small sticky grains requires destructive collisions which limit or even reverse the growth of the ``runaway" body.   So it remains to be shown whether this growth mechanism has the proper sign.

\subsection{Plastic Deformation}\label{plastic}
Car crashes show that plastic deformation can dissipate large amounts of kinetic energy in violent collisions.  This may not be directly relevant to planetesimal formation.  Only sufficiently ductile materials can permanently change shape in response to large stresses, others simply rupture.  The former category includes metallic elements and alloys, while the latter includes glass, bone, and crucially, rock.  Of course planetesimals must form prior to segregation of an iron-rich core, and evenly dispersed metallic elements probably cannot sustain plastic flow.  Would you feel safe driving a car that was $1/4$ metal and $3/4$ rock?

Even if planet forming material does deform plastically, we should be aware that the regime of stability is rather narrow.  Contact pressures only slightly above $Y$ give plastic flow that is confined by similarly large elastic stresses to a small region.  Somewhat larger stresses, $\sim 3 Y$, give ``uncontained" plastic flow that breaks through the surface allowing displaced material to escape (Johnson 1985).  During planetesimal formation, uncontained flow can cause mass loss by sub-sonic cratering and, at high enough energies, shattering, though more study of these outcomes is deserved.  Since confined plastic flow requires a narrow regime of contact pressures, its  relevance to the CCSH, which exposes solids to orders of magnitude variations in impact stresses, is not clear.

Plastic flow develops at an impact velocity, $v_Y$,  calculable from elastic Hertz theory (which ignores surface binding) by equating $K_\infty$ to the elastic deformation energy, $\int F d\delta$.  Requiring the maximum stress to be $Y$ gives :
\begin{equation}\label{yield}
\rho_{\rm s} v_Y^2 \simeq 26 Y^5/E^4,
\end{equation}
or $\simeq 10$ cm/s for medium soft steel.  Thus most collisions between metals do indeed involve plastic deformation (Johnson 1985, p. 361).
 
\subsection{Aggregate Restructuring and Viscoelastic Organics}\label{agg}
So far, we have implicitly assumed that collisions involve uniform, compact materials.  What if sticking relies on the detailed structure of impactors?  For instance porosity (i.e.\  ``fluffiness") should enhance dissipation.  However open structures are fragile, and moreover would be compacted by repeated impacts.
More plausibly, growing solids might be loosely bound ``rubble piles," i.e.\  an aggregate composed of many smaller units.  Restructuring of the sub-units during impacts could be quite dissipative.   
By analogy, collisions between bean bags (sacks filled with beans or foam pellets) have a lower coefficient of restitution 
than collisions between individual large beans or pellets.  Of course bean bags have a cloth cover that astrophysical rubble piles lack. 

Blum (1990) constructs a model along these lines by assigning low values of $E = 2 \times 10^4$ dyn/cm$^2$ and $Y = 100$  dyn/cm$^2$ to readily deformable aggregates.  A serious problem exists with models of this type.
Impact velocities, $v_{\rm coll} > v_Y \approx 10^{-3}$ cm/s exceed the yield stress, and the sound speed in this material is only $\sqrt{E/\rho_{\rm s}} \approx 2$ m/s.  Impacts in the cm/s --- 10 m/s range would be highly disruptive events, inducing ``uncontained" plastic flow and possibly driving shock waves into the material.  Blum ignores these possibilities and extrapolates results (e.g.\ for $\epsilon$) from the moderate speed regime, $\rho_{\rm s} v_{\rm coll}^2/Y \la 10^{-3}$, all the way to the hypervelocity regime, $\rho_{\rm s} v_{\rm coll}^2/Y > 10^{3}$, where they do not hold (Johnson 1985, p. 366).

These caveats aside, Blum's model gives a maximum sticking speed $0.5 < v_{\rm cr} < 5$ m/s for his ~ mm-sized aggregates, depending on impact parameter.  The predicted coagulation velocity limit for ices is quite low, $v_{\rm cr} \sim 3(R/{\rm mm})^{-1}~ {\rm mm/s}$ (as read from the graph).  Amazingly (since the included physics is quite different) this agrees very with equation (\ref{vcr}) as evaluated for ice, including the drop of $v_{\rm cr}$ with $R$.  Blum developed this model to explain experimental results which showed that mm-sized wax coated spheres (with a glass/wax mass ratio of 2 --- 5) have a $20 \pm 12 \%$ sticking probability for $v_{\infty} \approx 1~{\rm m/s}$.  It is a mystery how such results (low sticking probabilities, small sizes and moderate speeds) combined with the theoretically expected  decrease in $v_{\rm cr}$ with $R$ could be interpreted as supportive of the CCSH.  The sizable wax fraction invites skepticism, but it justified by the existence of organics in meteorite falls.

Along these lines, Kouchi et al.\ (2002) collided 5 mm copper spheres into targets with a mm-thick coating of organics, including PAHs and oxygen- and nitrogen-containing materials.  They report temperature-dependent sticking  at speeds up to $v_{\rm cr} \approx 5$ m/s at 250 K.  The threshold fell to  $v_{\rm cr}\la 2$ m/s and $v_{\rm cr} \ll 1$ m/s at 300 K and 200 K, respectively.  Kouchi et al.\ offer viscoelastic creep, a time-dependent relation between stress and strain, as an explanation for enhanced stickiness.  The role of plastic deformation in the copper is unclear.
The observed coagulation speeds are probably insufficient to make planetesimals,  in particular due to the expected decrease of $v_{\rm cr}$ with size.  Furthermore, the experimental conditions are not likely to be repeated in protostellar disks.  The targets had mm-thick layers of pure organics, while carbonaceous chondrites (the most carbon rich meteorites) have a bulk carbon content of only $\sim 3.5 \%$.

\section{Conclusion}\label{concl}

We studied the implications of particle velocity dispersions in protoplanetary disks  for the collisional growth of planetesimals.  Over a wide range of sizes, mm --- 10 m, sticking is strongly inhibited.  Available binding energies are negligible compared to kinetic energies, and a suitably efficient, yet non-destructive, dissipation mechanism has not been identified.  The most elegant solution to this problem is the Goldreich-Ward hypothesis, which uses collective gravity to avoid this problematic coagulation regime.

Particle concentration mechanisms, such as vortex trapping, secular instabilities, and quiescent nodes in turbulent flow (see YS for details) may play a role in planetesimal formation.  Higher densities favor gravitational instability, and could also promote collisional growth if collisional speeds are significantly reduced.  Regarding the possibility of long-lived particle-trapping vortices, only marginally coupled particles ($\tau_{\rm stop} \approx 1$) are efficiently trapped.  Smaller particles remain attached to streamlines and larger bodies can plow through the vortex.  This limits the overall reduction in relative velocities.

For many years, supporters of the Goldreich-Ward hypothesis were on the defensive due to valid concerns about the stirring provided by turbulence. Proposed solutions to this problem (Sekiya 1998, YS) should reestablish the viability of particulate gravitational instabilities.  To make the debate more productive, coagulation proponents should similarly acknowledge, and explain how  to overcome, the considerable energetic obstacles to sticking.

\acknowledgments 
I am indebted to my (proto)stellar role models: Frank, Chris, and Dave.  I thank Jeremy Goodman, Marc Kuchner, Scott Tremaine, and Bruce Draine for fruitful discussions.  This work was partly supported by  NASA Origins grant NAG5-1164.

\end{document}